\documentclass[a4paper,twocolumn,11pt,accepted=2023-05-12, colorlinks=true,hyperindex,allcolors=quantumviolet,breaklinks=true,,colorlinks=true,urlcolor=blue,citecolor=blue,linkcolor=blue]{quantumarticle}
\pdfoutput=1
\usepackage[utf8]{inputenc}
\usepackage[english]{babel}
\usepackage[T1]{fontenc}
\usepackage{amsmath}
\usepackage{hyperref}

\usepackage{tikz}
\usepackage{lipsum}

\pdfoutput=1

\usepackage[normalem]{ulem}
\usepackage{graphicx}
\usepackage{amsmath}
\usepackage{amssymb}
\usepackage{latexsym}
\usepackage{fancyhdr}
\usepackage{array}
\usepackage{amsfonts}
\usepackage{mathrsfs}
\usepackage{mathtools}
\usepackage{color}
\usepackage{verbatim}
\usepackage{physics}
\usepackage[caption=false]{subfig}
\usepackage{natbib}
\usepackage{enumitem}
\usepackage{tcolorbox}
\usepackage{bbold}
\begin{document}

\title{Relative Facts of  Relational Quantum Mechanics are
 Incompatible with Quantum Mechanics}

\author{Jay Lawrence}
\affiliation{Department of Physics and Astronomy,
Dartmouth College, Hanover, NH 03755, USA}

\author{Marcin Markiewicz}
\affiliation{International Centre for Theory of Quantum Technologies (ICTQT),
University of Gdansk, 80-308 Gdansk, Poland}

\author{Marek \.Zukowski}
\affiliation{International Centre for Theory of Quantum Technologies (ICTQT),
University of Gdansk, 80-308 Gdansk, Poland}

%\date{\today}

\begin{abstract}
Relational Quantum Mechanics (RQM) claims to be an interpretation of quantum theory [see \cite{Rovelli.21}, which  appears in the \textit{Oxford Handbook 
of the History of Interpretation of Quantum Physics}]. 
However, there are significant
departures from quantum theory: (i) in RQM measurement outcomes arise from interactions 
which entangle a system $S$ and an observer $A$ without decoherence,  and  (ii) such an 
outcome is a “fact” relative to the observer $A$, but it is not a fact relative
to another observer $B$ who has not interacted with $S$ or $A$ during the
foregoing measurement process. For $B$ the system $S \otimes A$ remains entangled. We derive a GHZ-like contradiction showing
that relative facts 
described by these statements are  incompatible with quantum theory.  Hence Relational Quantum Mechanics should not be considered an interpretation of
quantum theory, according to a criterion for interpretations that we have introduced. The criterion states that whenever an interpretation introduces a notion of outcomes, these outcomes, whatever they are, must follow the probability distribution specified by the Born rule.
\end{abstract}

\maketitle
\section{Introduction}

 Nearly a century after the birth of quantum theory, we have more different 
interpretations of it than ever, as well as a growing number of proposed 
alternative theories.  Many of these so-called interpretations are discussed and
grouped into categories by Cabello \cite{Cabello.17}. 
There are two broad categories (with subcategories that do not concern us), 
namely (i) “intrinsic realism”  - those which hold that “the probabilities of 
measurement outcomes are determined by intrinsic properties of the observed 
system,” and (ii) “participatory realism” - which hold that “quantum theory does 
not deal with intrinsic properties of the observed system, but with the experiences 
an observer or agent has of the observed system.”  One might also characterize 
these two categories with the adjectives “system-centric” and “observer-centric,” 
respectively.  Most of the current developments have occurred in the second 
category.  These include QBism, Rovelli’s Relational Quantum Mechanics 
and related ideas of Brukner, as well as the Copenhagen interpretation.  More
radical developments since the publication of Cabello’s paper incorporate ideas
of metaphysics, philosophy and psychology, together with interpretations of 
quantum theory \cite{Nurgalieva.20, Adlam.21, Cavalcanti.21, Brukner.22}.  
Discussions of these “interdisciplinary” developments would take us beyond 
the scope of this work.

In this paper we will focus on Relational Quantum Mechanics (RQM), which 
was introduced by Rovelli \cite{Rovelli.96}.   RQM is a body of 
ideas which has been extended with the help of a number of coworkers over the 
years.  It has helped to spawn variations by others, with subtle differences, 
but these works together represent an emerging theme within the 
observer-centric category.  Their ideas are illustrated by scenarios
involving two or more observers.  In the simplest case of two observers, one
observer ($A$) performs a measurement on a (typically simple) physical system
$S$ (say a qubit), and then the second observer $B$ measures the joint system 
consisting of $S$ and $A$. We shall refer to these and more complicated
versions as ``layered-observer'' scenarios.  One can view them as inspired by 
the well-known Wigner’s Friend scenario \cite{Wigner.61}, and an extension
by Deutsch \cite{Deutsch.85}.  These two earlier works had different motivations 
and drew conclusions unlike the current ones; they do not belong to the
current theme.  But, since they provide inspiration for current developments, 
we include brief descriptions in Appendix A for interested readers.

The current theme was initiated by Rovelli in his seminal work \cite{Rovelli.96}.
The ideas are illustrated by a scenario that introduces this work:  An observer
$A$ interacts with a system $S$ (say a single qubit), and realizes an outcome.  
The outcome is one of the possible values (an eigenvalue) of the 
measured observable.  This value is a {\it relative fact} for A.  It is 
{\it not} a relative fact for a second observer $B$ who has not interacted 
with $S$ or $A$ during the process.  RQM assumes that $B$ faces the situation 
in which $S$ and $A$ are in an entangled state representing a superposition of 
all possible outcomes.  Thus, according to RQM, B has a different, but equally 
valid account of the events as compared with $A$’s account. This is special to
RQM.  Common features of the above layered-observer theories involve
similarly unconventional ideas about measurement; for example, the rejection 
of the collapse postulate,\footnote{An observer-independent concept as understood 
in quantum mechanics (defined in the first paragraph of Sec. 2), as compared with the observer-dependent 
concept in RQM.} and measurement without decoherence - the embrace 
of relative facts.  RQM in particular holds that a measurement is simply an 
entangling interaction between {\it any} two systems;  it does not require 
a macroscopic measuring device.  We emphasize that when referring to 
RQM's description we use the term ``observer'' in this very broad sense: it can 
refer to any physical system interacting with another system. Such significant 
departures from quantum measurement theory create 
tension between these new theories and quantum mechanics. We will show how 
this tension manifests itself as a contradiction when it comes to predicted 
measurement outcomes.

General interest in layered-observer scenarios has grown since 
the appearance of the articles \cite{Frauchiger.18} and \cite{Brukner.18}.  
These draw inspiration from the extended Wigner’s Friend 
scenario of Deutsch \cite{Deutsch.85}, (Appendix A), which provided some 
intriguing background ideas.  Hence we will refer interchangeably to 
``extended Wigner-Friend’’ and ``layered-observer’’ scenarios.  Two of 
us proved a no-go theorem for extended Wigner-Friend scenarios involving  
relative outcomes for the Friend
\cite{Zukowski.21}, to counter the argumentation of  \cite{Frauchiger.18} 
and \cite{Brukner.18}, which use the no-collapse approach for some of the 
observers who play the role of Friends.

As the above examples show, there is a trend in the literature not to accept 
quantum measurement theory,  and instead claim  that one can have ``a unitary quantum mechanics'' without collapse, 
and importantly also  without the decoherence theory of measurement. In such 
theories, including RQM, measurement is an establishment of a system-device 
correlation (entanglement), and the introduction of `relative facts' as the  
measurement results with respect to an observer controlling the device.

The rejection of collapse ignores an essential aspect of the measurement 
postulates of quantum mechanics.  These postulates  entail the Born probability 
rule for measurement outcomes, with collapse as the post-measurement state-update 
rule.  The latter guarantees that a repeated measurement will produce the same 
result as the first.  In the initial 
formulations of quantum mechanics, it was presumed that collapse reflects the 
classical nature of macroscopic measuring devices, and the associated irreversible 
amplification process. It is now generally understood that collapse is described 
(and justified) by decoherence theory, in which the measuring apparatus is treated 
quantum mechanically, and assuming that its interactions with the system
and the environment are unitary. The interactions with the  environment cause
a rapid and irreversible decay of the off-diagonal elements of the density matrix 
(the coherences), leaving the diagonal elements as the Born rule probabilities of 
measurement outcomes, see e.g. \cite{Zurek.82}, \cite{Schlosshauer.04}, \cite{Schlosshauer.19}, \cite{ZUREK2022}.  
Thus, collapse manifests as a Bayesian conditional update.  A measurement cannot 
occur without the decoherence and irreversibility that inevitably accompany the 
action of a measuring apparatus, producing  a classical (probabilistic) description, 
see e.g. \cite{HAAKE}.

In this paper, we address specifically the Relational Interpretation, as 
anticipated in the concluding paragraph of  \cite{Zukowski.21}, and we focus on the 
central element of relative facts.   In common with \cite{Zukowski.21}, our proof 
is based on a three-qubit Greenberger-Horne-Zeilinger (GHZ) state: we exploit the 
perfect correlations predicted by quantum mechanics for measurement outcomes. 
These correlations impose constraints on the products of relative 
facts for two different observers ($A$ and $B$). 
   We will show that relative facts cannot satisfy these constraints, and cannot 
therefore be incorporated within quantum mechanics. In this way, Relational 
Quantum Mechanics is incompatible with quantum theory.  Therefore, according to
the following criterion, RQM should not be considered an interpretation
of quantum theory:
\begin{itemize}
    \item 
{\em Criterion for an interpretation:}  If an interpretation of quantum theory 
introduces some conceptualization of outcomes of a measurement, then  probabilities 
of these outcomes must follow the quantum predictions as given by the Born rule.
\end{itemize}
This includes any constraints, probabilistic or deterministic, implied by the Born 
rule.  Outcomes may include perceptions of observers.\footnote{Wigner 
and Deutsch refer to such outcomes in the scenarios described in Appendix A.} Our proof will show that RQM does not
satisfy this criterion.

We should emphasize that our proof in this work is based on the definition of 
relative facts within the RQM framework, and that it does not apply to relative 
facts as might be defined within the framework of QBism \cite{Pienaar.A.21}.  
The primary difference is that a relative fact in QBism refers to the statement 
of beliefs of an agent, and it is natural to expect differences between agents, 
which need not imply inconsistencies. In contrast, relative facts in Relational 
Quantum Mechanics refer to properties of a physical system $S$ with which 
$A$ interacts.

{\it Outline of the paper.} In Section 2, we review the postulates of RQM 
 and comment on the tension with quantum mechanics.  In 
Sec. 3, we lay out the ground rules and assumptions to be followed in the 
proof of the GHZ contradiction.  The proof itself is presented in Sec. 4.
It is self-contained; familiarity with the earlier GHZ-type reasonings is 
not assumed. In Section 5 (Further Remarks) we demonstrate the compatibility of 
mathematical steps in our proof with the rules of Relational Quantum Mechanics, 
and we argue that relative facts are a form of hidden variables. In Appendix A, 
we review the historical origins of extended Wigner-Friend scenarios, 
 and in B, we discuss a recent amendment to RQM.

\section{Tension between RQM and quantum mechanics}
 
Here we present the essential postulates of RQM of which we shall make use in our
arguments.  These will be accompanied by commentaries on the tension with 
 quantum mechanics. We refer here to the quantum theory which is used to 
predict quantum phenomena, which has not been falsified for the last 
98 years.  It is a theory which may be defined 
by the postulates as listed in standard textbooks. 
  %%%%%%%%%%%%%%%%%%%%%%%%%%%%%%%%%%%%%%%%%%%%%%%%%%%%%%%%%%%%%%%%%%%%%%%%%%%%%%%%%%%%%%%%%%%%%%%%%%%%
  \subsection{Principles of RQM (with commentary)}

Our description of RQM draws upon three articles, most prominently (i) Rovelli's 
entry in the Oxford Handbook of Quantum Interpretations \cite{Rovelli.21}, which 
sets out the main postulates of RQM.  The other two articles propose 
amendments to these main postulates.  The earlier of these is (ii) a reply of 
Di Biagio and Rovelli \cite{Biagio.22} to a recent critique by Pienaar 
\cite{Pienaar.21}, and the more recent is (iii) an article by Adlam and 
Rovelli \cite{Adlam.22}. 
  
We shall write the principles of RQM (marked by bullet points) with direct quotes 
of Rovelli where possible, and otherwise we shall paraphrase.

\begin{itemize}
    \item ``RQM interprets quantum mechanics as a theory about physical 
      {\it events}, or {\it facts}''.
    \item  ``A fact is quantitatively described by the value of a 
      variable or a set of variables.''
    %\item Facts are {\it sparse} [see below] and {\it relative}.
    \item ``Facts are ... realised {\it only} at the {\it interactions} 
      between (any) two physical systems.'' 
\end{itemize}
 Commentary:  According to the second and third bullet points, the realization 
of a fact in RQM corresponds to a measurement outcome, and this occurs without 
decoherence or the collapse of the state vector.  It conforms to the  
RQM idea that ``interaction is measurement.'' 
We will henceforth  refer to this concept as an ``RQM measurement''.
In  quantum mechanics, 
the interaction between two systems, the measured system and the ``pointer'' variable, {comprises}  what is called 
a ``premeasurement,'' which is a decoherence-free correlation between two systems
without an outcome. There is no known experiment in which  outcomes are obtained 
without the help of a macroscopic apparatus and without decoherence.

\begin{itemize}
    \item ``Facts are {\it relative} to systems that interact. That is, 
    they are {\it``labelled"} by the interacting systems.'' 
 \end{itemize}
This point is developed in the next two bullet points below, which we call {\em RQM postulates for Wigner-Friend scenarios}:
\begin{itemize}
    \item 
     ``...the Friend interacts with a system and a fact is realised {\em with respect 
     to the Friend}.''  But this fact is not realised with respect to Wigner, who was 
     not involved in the interaction...'' 
    \item 
    ``With respect to Wigner, it only corresponds to the establishment of an  
    'entanglement' ... between the Friend and the System.''
\end{itemize}
 Commentary:  These two bullet points conveniently summarize the ``active'' RQM principles that
we shall use in our proof in Sec. 4. They allow two different (but ``equally valid'', according to RQM) accounts of the events associated with the interaction between the Friend 
and the System.  In quantum mechanics, however, the Friend can be aware of 
an outcome only by reading the output of the apparatus. 
  
\begin{itemize}   
   \item
    ``Any system is an `observer:' a system with respect to which facts happen. 
    Decoherence characterizes observers with respect to which stable facts happen.''
\end{itemize} 
 Commentary: Quantum mechanics has a more restrictive concept of an
observer:  An observer is any object (animate or inanimate), which is capable of recording the (macroscopic and irreversible) result of a quantum measurement.  A microscopic
object, such as an atom belonging to a Bell pair, does not qualify as an observer.
There is no meaning 
to the statement that an atom ``realizes'' the value of its partner's spin component
$\sigma_z$.

\begin{itemize}
   \item The probability amplitude, $W(b,a)$,  determines the probability 
    for a fact $a$ given that a fact $b$ occurred.
   $W(b,a)$ determine probability amplitudes {\it only} if facts $a$ and $b$ 
   are relative {\it to the same systems}. 
    \item
    Assume that events $a$, $\{b_i\}$ and $c$ happen in a sequence, and that the 
    intermediate events within the set $\{b_i\}$ are mutually exclusive.
    If $c$ and $b_i$ have different ``labels'', that is, pertain to different systems, 
    the transition probability between $a$ and $c$ reads:
    \begin{equation} \label{RULE-1}
    P_{\textrm{unitary}}(c|a)=\left|\sum_iW(c,b_i)W(b_i,a)\right|^2.
    \end{equation}
\end{itemize}
 Commentary:  In our proof, all mathematical steps will be consistent with the above RQM statements about probability
amplitudes, as we shall confirm in Sec. 5.

\subsection{Amendments}

\noindent A first amendment to the principles quoted above was presented in \cite{Biagio.22}.  It is known as {\em the no-comparison rule}, 
\begin{itemize}
\item ``It is meaningless to compare events relative to different systems, 
unless this is done relative to a (possibly third) system.
 [...] comparisons can only be made by a (quantum–mechanical)
interaction. [...] Wigner [...] might compare the result of his measurement 
on S with that of F only by physically interacting with F in an appropriate
manner. There is no meaning in comparing facts relative to Wigner with 
facts relative to Friend, [...] apart from this direct physical interaction.''
  \end{itemize}
  A second amendment by \cite{Adlam.22} is intended to replace 
the ``first amendment'' above.

\begin{itemize}
 \item ``{\it cross-perspective links:}''
In a scenario where some observer Alice measures a variable V of a system S, 
then provided that Alice does not undergo any interactions which destroy the 
information about V stored in Alice’s physical variables, if Bob subsequently
measures the physical variable representing Alice’s information about the 
variable V, then Bob’s measurement result will match Alice’s measurement result.
\end{itemize}

\noindent Commentary:  The motivation for this amendment (see p. 3, 
paragraph 2 of \cite{Adlam.22}) is to ensure that a repeated measurement by 
another observer produces the same outcome (the state-update rule of quantum 
mechanics) or more specifically, that a measurement of Alice’s physical variable by Bob
will yield a result consistent with her measurement.   Further comments about 
cross-perspective links are given in Appendix B.

 We should emphasize that our no-go theorem for  relative facts of RQM in quantum theory does not require the use of the 
cross-perspectives links postulate.  However, if 
one assumes this postulate,
then our criterion for an interpretation (Section 1) together with its concrete implementation in the context of relative facts (Section 3) does not have to be invoked to show incompatibility between RQM and quantum mechanics.

%%%%%%%%%%%%%%%%%%%%%%%%%%%%%%%%%%%%%%%%%%%%%%%%%%%%%%%%%%%%%%%%%%%%%%%%%%%%%%%%%%%%%%%%%%%%%%%%%%%%%%%%%%%%%%%%%%%%%%%%%%%

\section{Strategy of Proof}

 There are many points of tension between RQM and quantum mechanics.  This tension
exists at the level of postulates or concepts which define the theory.  To prove 
that two theories, so defined, are actually different, or incompatible, we must work 
at the level of predicted measurement outcomes (as in our {\em criterion for an interpretation} in Sec. 1). We shall begin by considering a 
scenario built on the
Wigner-Friend postulates, which in turn are 
based on the properties of relative facts.  Eventually we will arrive at a contradiction 
between the predictions of RQM and quantum mechanics.  Specifically, we will 
show that relative facts predicted by RQM cannot take values that satisfy the 
constraints predicted by quantum mechanics for measurement outcomes. The tension
between RQM and quantum mechanics at the postulate level becomes a clear
incompatibility at the level of predicted outcomes.  

The scenario that we will construct is as follows:  The Friend ($A$) will make 
RQM measurements on a system $S$ consisting of three qubits in a GHZ state, 
thus acquiring three relative facts.  Following this, Wigner ($B$), who has 
interacted with neither $S$ nor $A$ during this measurement, will make RQM 
measurements on the compound system of $S$ and $A$, thus acquiring three 
relative facts of his own. The perfect correlations present in the initial
state of $S$ propagate through the evolution of the scenario and are 
manifested as constraints on the final outcomes, the relative facts of $A$ 
and of $B$. The final equations expressing these constraints are analogous
to those of Mermin in his proof \cite{Mermin.90} that noncontextual hidden 
variables are incompatible with quantum mechanics.  This may be of interest
to readers familiar with Mermin's work, but our proof is self-contained, and
familiarity with previous proofs is not assumed.

In the above scenario, we shall employ RQM measurements obeying the RQM rules 
of Sec. 2, most importantly the Wigner-Friend points.  We recall, with 
emphasis, that the term “RQM measurement” 
refers to the RQM concept (Sec. 2) of a measurement as an entangling 
interaction between two systems through which relative facts are realized. 
We use this term to avoid confusion with the concept 
of measurement in standard quantum measurement theory \cite{Zurek.82}, \cite{Schlosshauer.04}, \cite{Schlosshauer.19}, 
\cite{ZUREK2022},  
in which entangling interactions correspond only to pre-measurements (where decoherence does not occur),  but the completed measurements, with outcomes,  
{\em require} decoherence. RQM calls the latter results 
“stable facts”.

We shall also apply the rules of quantum mechanics, avoiding steps that
are explicitly forbidden by RQM  in its original version without the last amendment of \emph{cross-perspective links}, for the reasons discussed at the end of Section 2.  For example, we shall nowhere invoke 
the collapse postulate, and we shall nowhere compare the {\it values} of 
relative facts of different observers.  However, we will  implement our ``criterion for an interpretation'' by applying quantum 
mechanical constraints (the Born rule) to {\it products} of relative facts of different 
observers, by insisting that:
\begin{itemize}
    \item 
 Relative facts for different observers, who make different measurements, must take values that obey the constraints imposed by quantum mechanics. 
\end{itemize}

\noindent The above is motivated by the following reasoning 
concerning the logical structure of quantum theory.  Relative facts are a 
supplement to quantum mechanics.  RQM considers them to be outcomes (realizations) 
of RQM measurements.  If they do not satify the  constraints arising from the
Born rule probabilities, then RQM and quantum theory are incompatible with one 
another. And, according to our criterion, RQM is {\em not} an interpretation of 
quantum theory.

\section{Proof of GHZ Contradiction}
 
In the next three subsections we shall first define the prepared state of the
system $S$; next, the RQM measurements on $S$ by the observer A; and lastly the 
RQM measurements on the compound system $S \otimes A$ by the observer $B$.
In a fourth subsection we shall identify the perfect GHZ correlations, and in
subsection five we complete the proof.

\subsection{The System S}

\begin{figure}
    \centering
    \includegraphics[width=0.9 \columnwidth]{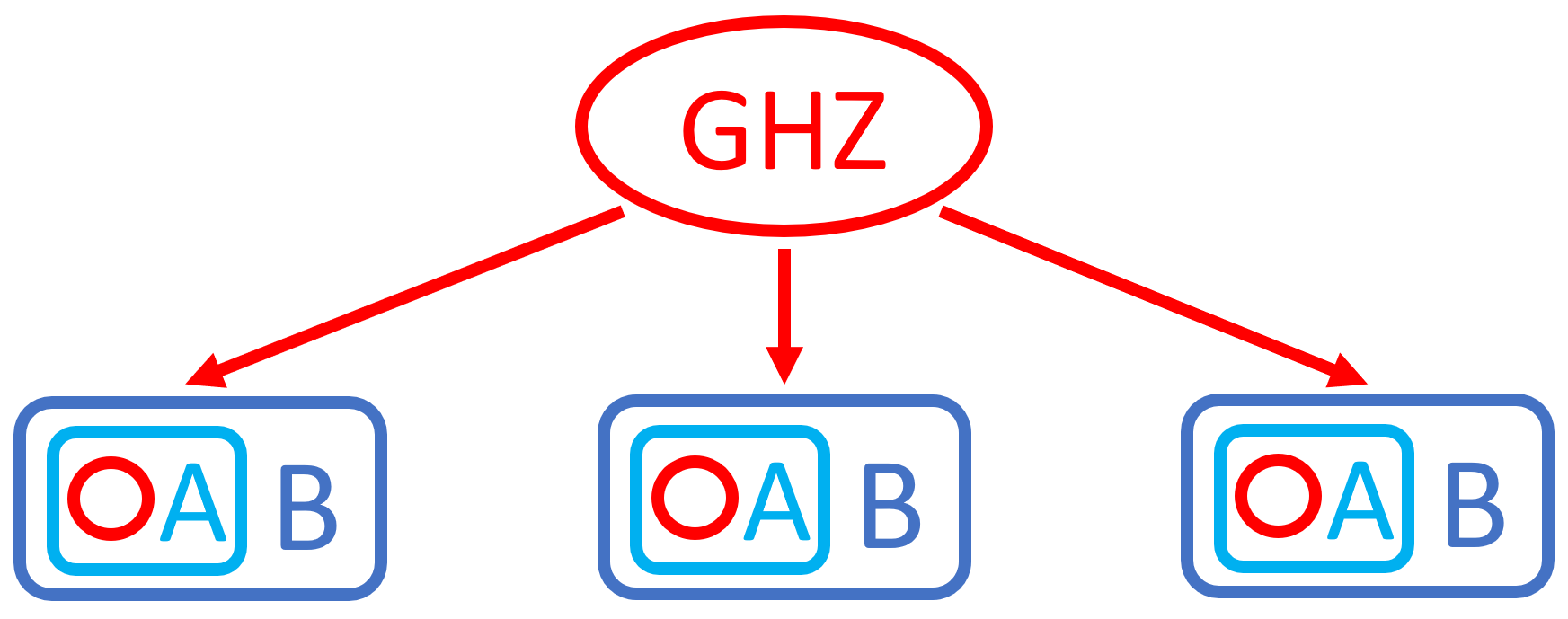}
    \label{figGHZ}
    \caption{Three qubits are prepared in a GHZ state.  Observer $A$ then
    performs separate measurements on each qubit $S_m$, $m=1,2,3.$ Finally 
    observer $B$ performs separate measurements on each pair $S_m\otimes A_m$, 
    which together comprise the compound object $S \otimes A$. {\em The arrows 
    and separations in the figure do not depict spatial separation of
    subsystems:} they merely emphasize that  measurements are performed on the
    individual objects (labeled by $m$).}
\end{figure}

%%%%%%%%%%%%%%%%%%%%%%%%%%%%

Our system $S$ consists of three qubits ($S = S_1\otimes S_2\otimes S_3)$ in 
a GHZ state, as pictured in Fig. 1. We shall write this state below, but 
first we introduce our notation for the three complementary (one-qubit) bases.
They will be referred to later in  descriptions of the anticipated
measurements by $A$ and $B$, as well as the initial state itself.

The basis states are denoted by
$\ket{\pm1^{(n)}}_{S_m}$, where $\pm 1$  are the state indices 
(in any basis), $n=1,2,3,$ denote the bases, and $m$ specifies the qubit. The 
bases $n=2,3$ expressed in terms of the  standard basis $n=1$ are  
\begin{eqnarray} \label{BASIS2}
\ket{\pm1^{(2)}}_{S_m}=\frac{1}{\sqrt{2}}\left(\ket{+1^{(1)}}_{S_m}
\pm \ket{-1^{(1)}}_{S_m}\right),\nonumber\\
\end{eqnarray}
and 
\begin{eqnarray} \label{BASIS3}
\ket{\pm1^{(3)}}_{S_m}=\frac{1}{\sqrt{2}}
\left(\ket{+1^{(1)}}_{S_m} \pm i \ket{-1^{(1)}}_{S_m}\right).\nonumber\\
\label{basis3} 
\end{eqnarray}
These bases $n=1,2,3$, form the standard trio of mutually complementary bases, 
typically called $Z,X,Y$, respectively. Following convention for qubit states, 
$l=\pm 1$ are the eigenvalues of the associated Pauli operator ($\sigma_n$)
in the states  $\ket{l^{(n)}}_{S_m}$.
%The general notation for the basis states will be  $\ket{l^{(n)}}$, with the 
%superscript $(n)$ labeling specific bases.
    
We choose a GHZ state as the initial state of $S$, 
\begin{eqnarray}
&\ket{GHZ}_{S_1S_2S_3}&\nonumber \\&=
  \frac{1}{\sqrt{2}}\left(\otimes_{m=1}^3  \ket{+1^{(1)}}_{S_m}
    +\otimes_{m=1}^3  \ket{-1^{(1)}}_{S_m} \right),&\nonumber\\
\label{GHZ1}
\end{eqnarray}
involving standard basis states for each qubit.

\subsection{RQM measurements by A}
    
Observer $A$ now makes RQM measurements separately on each qubit $S_m$, interacting 
 so as to perform each measurement in the basis $n=3$.  According to RQM, each 
measurement then consists of a unitary entangling evolution,  $\hat U_m^{SA}$, 
defined by:
\begin{eqnarray} \label{MEAS-F}
    \hat U_m^{SA}\left(\ket{l^{(3)}}_{S_m}\ket{\textrm{initial}}_{A_m}\right)&=& 
    \ket{l^{(3)}}_{S_m}\ket{l^{(3)}}_{A_m}\nonumber\\
    &\equiv&
    \ket{l^{(3)}}_{SA_m},
    \label{SFbasis3}
\end{eqnarray} 
where we have introduced shorthand notation $SA_m$ for $S_m \otimes A_m$. The three 
interactions together comprise an RQM measurement on the system $S$ in the product 
basis $\otimes_{m=1}^3\ket{l_{m}^{(3)}}_{S_m}$. 
  
To understand the resulting entangled state of $S \otimes A$ in a transparent manner,
we first expand the initial GHZ state (of just the $S_m$) in the $n=3$ basis:
\begin{eqnarray}
     \ket{GHZ}_{S_1S_2S_3}
    =\sum_{p,q,r=\pm1} c^{333}_{pqr}\ket{p^{(3)}}_{S_1}\ket{q^{(3)}}_{S_2}
    \ket{r^{(3)}}_{S_3}.\nonumber\\
    \label{pqr1}
\end{eqnarray} 
The expansion coefficients are clearly given by
\begin{equation}
\label{Cn3}
    c^{333}_{pqr} = \frac{1}{\sqrt 2}\sum_{l=0}^1 \langle p^{(3)}|l^{(1)}\rangle 
\langle q^{(3)}| l^{(1)}\rangle \langle r^{(3)}
|l^{(1)}\rangle,
\end{equation}
where subscripts in $c^{333}_{pqr}$ are basis state indices, and superscripts are
the respective bases they refer to. Combining the three 
entangling operations (\ref{MEAS-F}) with (\ref{pqr1}), we have 
\begin{eqnarray} \ket{GHZ}_{SA}&=&\otimes_{m=1}^3{\hat U_m^{SA}}\left(\ket{GHZ}_{S}
      \otimes_{j=1}^3 \ket{\textrm{initial}}_{A_j}\right) \nonumber \\
      &=&\sum_{pqr=\pm1} c^{333}_{pqr}\ket{p^{(3)}}_{SA_1}
     \ket{q^{(3)}}_{SA_2}\ket{r^{(3)}}_{SA_3}.\nonumber\\
     \label{pqr2}
\end{eqnarray} 
     According to RQM, observer A realizes a trio of relative facts -- 
    specific values ($\pm 1$) for each of $p$, $q$, and $r$, all referring to basis 
    $n=3$.  We shall refer to these as $\mathcal A_1$, $\mathcal A_2$, and $\mathcal A_3$, respectively, for later 
    reference.  According to RQM, as mentioned earlier, these are {\it not} relative 
    facts for B, who has not interacted with $S$ or $A$ during the above process. We
    shall discuss $B$'s role in the next subsection, but let us first make a couple 
    of observations.
    
    The equality of coefficients in Eqs. (\ref{pqr1}) and (\ref{pqr2}) shows the exact 
    isomorphism between GHZ states of the $S$ and $S \otimes A$ systems.  From this we
    may conclude that, although the system $S \otimes A$ has a larger Hilbert space, it 
    accesses an {\it effective} Hilbert space of the same dimension (eight) as that of 
    the three-qubit system $S$. This means that the number of possible independent  
    measurement outcomes with nonzero probability, as inferred from either equation, 
    is the same. 
    
    We should also note here that there are no correlations between measurement 
    outcomes in the $n=3$ basis.  One can see this by showing that all possible 
    measurement outcomes will occur with equal probability; that is, all of the 
    coefficients in the superposition (\ref{Cn3}) have $|c_{pqr}^{333}|^2 = 
    \frac{1}{8}$.  Correlations will appear with measurements taken in the $n=2$ 
    basis.
     
     \subsection{{RQM} measurement by $B$ on the system $S \otimes A$}

     Now $B$ interacts with each of the subsystems $S_mA_m$ so as to perform RQM 
     measurements in the basis $n=2$.  Note that this basis (and the others) are defined in
     the effective (eight-dimensional) Hilbert space occupied by the state  $\ket{GHZ}_{SA}$ \eqref{pqr2}, as described two paragraphs above. 
     RQM assigns this state to the system $S \otimes A$ now faced by $B$.  
     To explore the consequence of $B$'s interactions,  
     we re-express $\ket{GHZ}_{SA}$ in terms of the $n=2$ basis for 
     $S_m\otimes A_m$, whose states are 
     related to those of the $n=3$ basis, $\ket{\pm 1^{(3)}}_{SA_m}$  , by:
     %%%%%%%%%%%%%%%%%%%%%%%%%%%%%%%%%%%%
\begin{eqnarray}  
\ket{\pm 1^{(2)}}_{SA_m} \equiv \frac{1}{\sqrt{2}} 
\bigg( \ket{+1^{(3)}}_{SA_m} \pm i \ket{-1^{(3)}}_{SA_m} \bigg)\nonumber\\
\label{SFbasis2}
\end{eqnarray}
modulo an irrelevant overall phase factor.  $B$'s interactions with
$SA$ may now be expressed by
unitary transformations $\hat U^{SAB}_m$ which have the familiar entangling features: 
 \begin{eqnarray}
   \hat U^{SAB}_m\left( \ket{l^{(2)}}_{SA_m}\ket{\textrm{initial}}_{B_m}\right)
    &=& \ket{l^{(2)}}_{SA_m}\ket{l^{(2)}}_{B_m} \nonumber\\ 
    &\equiv&
    \ket{l^{(2)}}_{SAB_m},
    \label{SBbasis3}
\end{eqnarray} 
where $B_m$ denotes the subsystem of $B$ which interacts with $SA_m$, and $SAB_m$ 
stands for $S_m\otimes A_m \otimes B_m$.  The state of the system $SAB$ after 
these interactions reads:
\begin{eqnarray}
     &\ket{GHZ}_{SAB} &\nonumber \\
    &=\sum_{pqr} c^{222}_{pqr}\ket{p^{(2)}}_{SAB_1}\ket{q^{(2)}}_{SAB_2}
    \ket{r^{(2)}}_{SAB_3}.&\nonumber\\
    \label{GHZ-222-AB}
\end{eqnarray}
 According to RQM, the entangling interactions comprise a measurement, 
and B realizes relative facts - specific values ($\pm 1$) for each of $p$, $q$, and 
$r$, but now referred to basis $n=2$.  We call these relative facts $\mathcal B_1$, $\mathcal B_2$, 
and $\mathcal B_3$ for later reference.  At this stage, the observers $A$ and $B$ have 
completed all RQM operations required for this gedanken experiment. We can now 
consider correlations between different possible outcomes.

\subsection{GHZ correlations}

For the sake of transparency, before considering measurement outcomes per se, we 
discuss correlations inherent in the state $\ket{GHZ}_{SAB}$ (\ref{GHZ-222-AB})
itself, as manifested in the expansion coefficients $c_{pqr}$.  As a first example, 
we note that $\ket{GHZ}_{SAB}$ is isomorphic with $\ket{GHZ}_S$ when expressed in 
the basis $n=2$ for all qubits. Hence one can easily show that the amplitudes 
$c^{222}_{pqr} \neq 0$ only if:
\begin{equation}p^{(2)}q^{(2)}r^{(2)}=1.
\label{C222} \end{equation}
This represents one of the four perfect GHZ correlations, which impose constraints
on products of the individual eigenvalues. The analog of Eq. (\ref{Cn3}) shows that 
the three individual measurement outcomes are random, but the product of all three 
is deterministic (and in this case, equal to unity).  

We shall identify three more perfect correlations, which will impose similar 
constraints on other combinations of measurement outcomes already obtained by 
$A$ and $B$, without requiring any further measurements.  We can derive these 
additional constraints as follows.  Unitary transformations like \eqref{SBbasis3} 
can be put in a form explicitly acting  on all three parts $SAB_m$. One can 
introduce $\hat {\textbf{U}}^{SAB}_{(m)}=\hat U^{SAB}_m \otimes \hat I_{nk}$, 
where $\hat I_{nk}$ is a unit operator acting on systems $S_{n}AB\otimes S_{k}AB$, 
where $m,n,k=1,2,3$ and cyclic permutations of these numbers. Note that the 
three transformations $\hat {\textbf{U}}^{SAB}_{(m)}$ ($m=1,2,3$) mutually 
commute, and thus their order of application is immaterial. 

We can now find the additional constraints on eigenvalues ($p,q,r$) by applying  
each of the three operations to the state $\ket{GHZ}_{SA}$.  Starting with 
$\hat {\textbf{U}}^{SAB}_{(1)}$, one generates the following Hilbert space vector,
\begin{eqnarray}
     &&\ket{GHZ}_{(SAB_1)(SA_2)(SA_3)}\nonumber\\
     &&=\hat{\textbf U}^{SAB}_{(1)}\left(\ket{GHZ}_{SA}\otimes_{j=1}^3
     \ket{\textrm{initial}}_{B_j}\right)\nonumber \\
    &&=\left(\sum_{pqr} c^{233}_{pqr}\ket{p^{(2)}}_{SAB_1}\ket{q^{(3)}}_{SA_2}
    \ket{r^{(3)}}_{SA_3}\right)\nonumber\\
    & &\otimes_{j=2}^3\ket{\textrm{initial}}_{B_j}.
    \label{GHZ-233-BAA}
\end{eqnarray}
Exploiting the isomorphism between this state and $\ket{GHZ}_{S_1S_2S_3}$ when
expressed in bases $n=2,3,3,$ respectively, one can show that 
$c^{233}_{pqr} \neq 0$ \textit{only if}:
\begin{equation}p^{(2)}q^{(3)}r^{(3)}=-1.
\label{C233} \end{equation}
This is the second perfect GHZ correlation, written as a state-imposed constraint 
on combinations of individual eigenvalues.  The third perfect correlation arises 
from the application of $\hat{\textbf U}^{SAB}_{(2)}$ to  $\ket{GHZ}_{SA}$. The 
resulting Hilbert space vector is $\ket{GHZ}_{(SA_1)(SAB_2)(SA_3)}$.  This has 
non-zero amplitudes for 
\begin{equation}p^{(3)}q^{(2)}r^{(3)}=-1
\label{C323}. \end{equation} 
Finally, $\hat{\textbf U}^{SAB}_{(3)}$
generates the Hilbert space vector $\ket{GHZ}_{(SA_1)(SA_2)(SAB_3)}$, with
non-zero amplitudes for
\begin{equation}p^{(3)}q^{(3)}r^{(2)}=-1.
\label{C332} \end{equation}
Thus we have four constraints,
Eqs \eqref{C222}, \eqref{C233}, \eqref{C323}, and \eqref{C332}, on products of 
individual eigenvalues that characterize the GHZ state (\ref{GHZ-222-AB}), and 
others related by unitary transformations.
 
 \subsection{GHZ contradiction with relative facts}

 The GHZ states are superpositions which allow all combinations 
of individual eigenvalues whose products satisfy the constraints \eqref{C222}, \eqref{C233}, \eqref{C323}, and \eqref{C332}.  The output of each 
set of RQM measurements is just one of these  possibilities, forming a trio 
of relative facts: First, $A$'s measurement on $\ket{GHZ}_{SA}$ 
(Eq. \ref{pqr2}), results in $\{p^{(2)}, q^{(2)},r^{(2)}\} \rightarrow 
\{\mathcal A_1,\mathcal A_2,\mathcal A_3\}$, a specific set of facts relative 
to observer $A$ (but not to $B$).  Second, $B$'s measurement on $\ket{GHZ}_{SAB}$ 
(Eq. \ref{GHZ-222-AB}), results in $\{p^{(3)}, q^{(3)},r^{(3)}\} \rightarrow 
\{\mathcal B_1,\mathcal B_2,\mathcal B_3\}$, a specific set of facts relative to 
observer $B$ (but not to $A$).  It is known to both observers that 
 two sets of measurements were performed resulting in six relative
facts,\footnote{This conforms to the example of the
Wigner-Friend scenario described by Rovelli in
(Rovelli, 2021)} even though only three of these are known to Alice (according to RQM), and only the other three are known to Bob.

With this, we can implement our technical criterion, as stated in the bullet point in Sec. 3:   the relative facts, 
$\{\mathcal A_1,\mathcal A_2,\mathcal A_3, \mathcal B_1, \mathcal B_2,\mathcal B_3\}$, 
whatever these may be, must follow the predictions of quantum theory, irrespective 
of whether they are realized with respect to the same or different observers, 
who make measurements in the same or different bases.

Thus, rewriting the GHZ constraints of Eqs. \eqref{C222}, \eqref{C233}, 
\eqref{C323}, and \eqref{C332} in terms of relative facts, we have
\begin{eqnarray}
\label{constraints}
\mathcal B_1\mathcal B_2\mathcal B_3&=&+1, \nonumber \\
\mathcal B_1\mathcal A_2\mathcal A_3&=&-1,\nonumber \\
\mathcal A_1\mathcal B_2\mathcal A_3&=&-1,\nonumber \\
\mathcal A_1\mathcal A_2\mathcal B_3&=&-1. 
\end{eqnarray}
The combination of all four equations leads to
\begin{equation} \label{allsquared} 
    \mathcal A_1^2 \mathcal A_2^2 \mathcal A_3^2 \mathcal B_1^2 \mathcal B_2^2 
    \mathcal B_3^2 = -1.
\end{equation}
This means that there is no set of real solutions $\{\mathcal A_m\}$ for which 
a set of real solutions $\{\mathcal B_m\}$ exists, and its converse.  The 
logical possibilities are:
\begin{itemize}
    \item If we allow that the relative facts $\{\mathcal A_m\}$ are
realized by the observer A, then Bob’s measurements cannot
produce the relative facts $\{\mathcal B_m\}$ satisfying the constraints
[\ref{constraints}] imposed by quantum mechanics.
    \item If we assume, on the other hand, that B’s measurements 
produce relative facts $\{\mathcal B_m\}$, then the relative facts
$\{\mathcal A_m\}$ do not exist, and in fact never existed.  For if they 
had, then B could not have realized the required $\{\mathcal B_m\}$. 
\end{itemize}

\noindent In either case, relative facts predicted by RQM lead to contradictions 
with the predictions of quantum mechanics based on GHZ correlations.  Therefore, 
the concept of relative facts cannot be accommodated within quantum mechanics 
without rendering it internally inconsistent; Relational Quantum Mechanics is 
incompatible with quantum mechanics. 
Since the GHZ correlations follow from the Born rule, Relational Quantum Mechanics, according to our 
{\em criterion}, cannot be considered an
interpretation of quantum mechanics. 

 If one additionally invokes the cross-perspective links postulate, then results $\mathcal A_i$ are effectively assumed to exist, and are merely unknown to Bob (see Appendix B), and thus we have a contradiction without invoking our technical criterion from Sec. 3, which is not a part of RQM. Therefore in such a case the contradiction found by us is an internal contradiction within RQM (augmented with cross-perspective
links postulate).

\section{Further Remarks}

\subsection{ Consistency with the RQM assumptions concerning probability amplitudes}

 Note that the constraint (\ref{C233}) is in full concurrence with the RQM rule 
 (\ref{RULE-1}). In the case of quantum mechanics it stems from: 
\begin{eqnarray}
\label{Prob233GHZ}
&&P\left(\left(p^{(2)},q^{(3)},r^{(3)}\right)\Big|GHZ\right)\nonumber\\
&&=\left|\braket{p^{(2)},q^{(3)},r^{(3)}}{GHZ}_S\right|^2,
\end{eqnarray} 
which holds due to the isomorphism between the final state \eqref{GHZ-233-BAA} and 
the ordinary GHZ state expressed in respective local bases.  In RQM the transition 
probability $P\left(\left(p^{(2)},q^{(3)},r^{(3)}\right)\big|GHZ\right)$ applies on 
the ``initial'' side to the three qubits, and on the ``outcomes'' (relative values) side 
to $B_1$, $A_2$ and $A_3$ subsystems respectively. The relative facts 
$\left(p^{(2)},q^{(3)},r^{(3)}\right)$ have different RQM labels (which means that 
they are realised with respect to different systems). The formula for the quantum 
amplitude appearing in the right-hand-side of  \eqref{Prob233GHZ}  can be re-expressed 
in terms of the RQM amplitudes in accordance with the rule \eqref{RULE-1}:
\begin{eqnarray}
&&\braket{p^{(2)},q^{(3)},r^{(3)}}{GHZ}_S=c^{233}_{pqr}\nonumber \\
&&=\sum_{p^{(3)}=\pm1}w\left(p^{(2)}, p^{(3)}\right) w
\left(\left(p^{(3)},q^{(3)},r^{(3)}\right), GHZ\right).\nonumber\\
\end{eqnarray}
On the one hand each of the RQM amplitudes applies to different systems in the 
\textit{in} and \textit{out} sides, e.g. in $w\left(p^{(2)}, p^{(3)}\right)$, 
$p^{(2)}$ is a relative fact about system $S\otimes A$ realised with respect to 
the observer $B$, whereas relative fact $p^{(3)}$ is about the initial system $S$   
realised with respect to the observing system $A$.  On the other hand the values of 
the RQM amplitudes, put here generally as $w(y,x)$, are due to isomorphisms between 
the final states after entangling interactions and the initial GHZ state, formulas 
(\ref{GHZ1}-\ref{GHZ-233-BAA}), given by  the respective quantum amplitudes 
$\braket{x}{y}_S$ calculated for the qubits $S=S_1\otimes S_2 \otimes S_3$, that 
is $w(x,y)=\braket{x}{y}_S$.

\subsection{Relation with no-go theorem for non-contextual hidden variables}

The four equations (\ref{constraints}) are identical to those written by Mermin 
[Equations (6) in \cite{Mermin.90}] in his refutation of hidden variables 
theories of either local or non-contextual character.  His proof is based on a 
GHZ state of three spins [our Eq. \eqref{GHZ1}] and generalizes the original 
theorem of Greenberger, Horne, and Zeilinger (GHZ) \cite{GHZ.89}. 

We should stress that the analogy with Mermin's proof is only mathematical, and not 
physical. Apart from the three-qubit GHZ states, the physical  situations and scenarios
are different.
In Mermin's  
no-go for non-contextuality proof there is one observer, 
and the reasoning involves four separate experiments.  In
our case, two observers perform a single experiment. In Mermin's case, the GHZ
contradiction arises only when one explicitly assigns 
non-contextual 
character to the variables in the equations.\footnote{In the version of Mermin's proof involving local hidden variables, there are three spatially separated observers, each choosing between two local measurements.}
In our case, we follow RQM, in which
no hidden variables exist initially. However, they come into existence when Alice
interacts with $S$ and realizes relative facts.  These relative facts are, 
effectively, hidden variables. 

Expanding on the above,
Alice's relative facts are hidden 
variables according to the following
definition:  A hidden variable is any 
additional notion, or feature of a
physical variable, parameter or state,
which is present in a given interpretation and does not appear in
the quantum mechanical formalism.
Such variables are more subtle in 
the Relational Interpretation: {\em they are produced in interactions between systems, not during preparation 
procedure of the initial quantum state}.

\subsection{Final note} 

 If relative facts were to have any meaning \emph{in} quantum theory,
they would have to  satisfy the  constraints that could
emerge from the quantum mechanical description of any experimental situation.  We have shown that this is impossible.

As a concluding remark we provide the following quotation from Wheeler \cite{WHEELER}:
\begin{itemize}
 \item
For a proper
way of speaking we recall once more that it
makes no sense to talk of the phenomenon
until it has been brought to a close by an
irreversible act of amplification: “No elementary phenomenon is a phenomenon until it is
a registered (observed) phenomenon.”
\end{itemize}
See also \cite{WHEELER-2}, and a review \cite{RMP-DELAYED}.

\section*{Acknowledgements}
MM and MZ are  supported by  Foundation for Polish Science (FNP), IRAP project ICTQT, contract no. 2018/MAB/5, co-financed by EU  Smart Growth Operational Programme.
MZ thanks Michail Skotiniotis for a discussion on the link of all that with non-contextual hidden variables. MZ also thanks the late Jarek Pykacz for years of discussions on the subject of hidden variables.

%apsrmp4-2.bst 2018-12-27 (MD) hand-edited version of apsrmp4-1.bst
%Control: author (3) reversed first dotless

\newpage
\appendix
\section{The First Wigner's Friend Scenarios}
\label{app:A}

\centerline{{\bf Wigner (1963)}}
\medskip

E. P. Wigner eloquently expressed the view that the state vector 
of a quantum system is a statement of an observer's knowledge 
about the system, based on the available information.  At the 
same time he followed {} quantum theory, including the 
collapse postulate. With the goal of locating the site of 
collapse in the von Neumann measurement chain \cite{vN.32}, he 
introduced his Friend in 1963 \cite{Wigner.61}, and placed her in the 
chain between the measurement apparatus and Wigner himself. The
Friend informed Wigner that she had observed a definite outcome,
creating a tension with Wigner's expectation, based on unitary
evolution, that one should instead find a superposition. This is 
the so-called ``Wigner's Friend Paradox,'' although for Wigner it 
was not a paradox:  For Wigner, it formed the argument that  
the ``collapse'' occurred in the consciousness of his Friend.  
Nevertheless, the ``paradox'' is a prominent theme in the 
current literature.

\bigskip
\centerline{{\bf Deutsch (1985)}}
\medskip

Deutsch introduced a more elaborate scenario in 1985 \cite{Deutsch.85} that has inspired some of the current variations, 
to different purposes.  Deutsch's motivation was to demonstrate 
the possibility, in principle, of distinguishing between the 
Copenhagen and Many-Worlds interpretations.  To this end, he 
believed it to be necessary for one observer (say Wigner) to perform 
measurements on a compound system containing another observer 
(say Friend), and the spin of an atom which Friend has measured.
Admittedly, the thought experiment is beyond experimental 
capabilities at present, if not forever. Nevertheless, here 
is a very brief outline:

The Friend measures the spin of the atom and reports to Wigner 
that she has observed a definite outcome, but does not reveal 
which outcome.  In a radical departure from the previous 
scenario, Wigner now performs an experiment on the compound 
system of the spin and the Friend, $S \otimes A$, including 
the consciousness of the Friend.  The experiment involves the 
reversal of the Hamiltonian dynamics of the Friend's 
consciousness, the atom's trajectories, and the spin detection 
apparatus - culminating finally in a remeasurement of the atom's spin.  A pure state indicates Many-Worlds evolution, whereas a mixed state indicates Copenhagen-like collapse.

It is important to stress a couple of points in 
assessing Deutsch's influence on later work.  First, the 
Friend's spin measurement is described only vaguely. There 
is a reference to ``sense organs,'' which represent some 
unspecified combination of the atom's detection apparatus and 
the Friend's consciousness.  Nonetheless, it is clear that 
the measurement process is purely unitary.  Second, Wigner's
expectation of a superposition state of $S \otimes A$ is also 
based on his (momentary) assumption of unitary evoution. Both 
assumptions are consistent with the Many-Worlds framework 
assumed by Deutsch.  However, these assumptions (of unitarity) 
are carried over into later scenarios in treatments that do not admit collapse into their descriptions, while
retaining a one-world perspective. This is the source of a number of inconsistencies 
in later works, see e.g. \cite{Frauchiger.18} and \cite{Brukner.18}, and the critique in \cite{Zukowski.21}.

\section{Cross-perspective links}

The \textit{cross-perspective links} postulate clearly defines the prescription for Bob to get information about Alice's result. The method is deterministic. Alice's relative fact determines Bob's outcome in this prescription, despite the other postulate of RQM which says that he faces the situation in which S and A are entangled. 
Thus, Alice's relative fact is effectively a hidden variable for Bob.

This postulate is aimed at establishing in RQM the repeatability of outcomes of measurements by any observer who subsequently measures in the same basis. It creates an obvious tension with the postulate that after Alice's interaction/RQM-measurement Bob faces an entangled state of the system with Alice, with the Schmidt basis for $S \otimes A$ being the measurement basis of $A$.
The introduction of {\em cross-perspective links}  clarifies the status of  Alice's relative facts. 
Namely, if we assume that for every two observers A and B, there exists a precisely defined operational way for B to {\em deterministically} recover outcomes of a measurement done by A, 
this means that A's outcomes have definite values, which are in principle accessible to the other observers, but without a direct {\em specific} interaction are just unknown to them. However, since the process 
of revealing them is fully deterministic, as effectively assumed by  {\em cross-perspective links}, this lack of knowledge is fully ``classical'', and has nothing to do with quantum randomness, superpositions etc. 
Hence, it is now clear that  axioms of RQM extended with  {\em cross-perspective links} imply that  every observer has access to his/her outcomes, and a deterministic procedure to get to know outcomes of any other observers, which means that in fact all outcomes exist for all observers, but some of them are just unknown.   This conclusion is supported by
the statement in \cite{Adlam.22}, p. 8, first paragraph,  which reads as
follows:  ``...{\it with the addition of the postulate of cross-perspective
links it no longer seems possible to insist that everything is relational -- or at least,
it is no longer necessary to do so -- because this postulate implies that the information stored in
Alice’s physical variables about the variable V 
of the system S is accessible in principle to any
observer who measures her in the right basis, so 
at least at an emergent level this information
about V is an observer-independent fact. This 
suggests that the set of ‘quantum events’ should
be regarded as absolute, observer-independent 
features of reality in RQM, although quantum
states remain purely relational. Thus we continue 
to endorse the sparse-flash ontology for RQM
as advocated in refs [15,16]: however we now regard the pointlike quantum events or ‘flashes’ as
absolute, observer-independent facts about reality, rather than relativizing them to an observer}.''

In summary, the  meaning of the new postulate is that RQM, when addressing the results of A in relation to observer B, gives to the results the status of existence. They are merely ``unknown'' to B, being accessible by measurement as dictated in {\em cross-perspective links}.

\end{document}